\documentclass[11pt]{article}
\usepackage{xspace}
\usepackage{graphicx}
\usepackage{amsmath}
\usepackage{amssymb}
\usepackage{color}

\definecolor{Red}{rgb}{1,0,0}
\definecolor{Green}{rgb}{0,1,0}
\definecolor{Blue}{rgb}{0,0,1}
\definecolor{Black}{rgb}{0,0,0}

\def\beq{\begin{equation}}
\def\eeq#1{\label{#1}\end{equation}}
\def\eeqn{\end{equation}}

\def\beqa{\begin{eqnarray}}
\def\eeqa#1{\label{#1}\end{eqnarray}}
\def\eeqan{\end{eqnarray}}

\let\bar=\overbar

\def\Dslash{\not{\hbox{\kern-4pt $D$}}}
\def\dslash{\not{\hbox{\kern-2pt $\del$}}}

\def\msb{{\bar{\ssstyle M \kern -1pt S}}}

\def\Title#1{\begin{center} {\Large {\bf #1} } \end{center}}

\begin{document}

\Title{The State of the Art of \\ Neutrino Cross Section Measurements}

\bigskip\bigskip


\begin{raggedright}  

{\it Deborah A. Harris\index{Harris, D.A.},\\
Fermi National Accelerator Laboratory\\
Batavia, Illinois 60510 USA}\\

\end{raggedright}
\vspace{1.cm}

{\small
\begin{flushleft}
\emph{To appear in the proceedings of the Prospects in Neutrino Physics Conference, 15 -- 17 December, 2014, held at Queen Mary University of London, UK.}
\end{flushleft}
}

\section{Introduction}

The study of neutrino interactions has recently experienced a renaissance based on its importance in measuring neutrino oscillations.  Neutrino oscillation experiments depend critically on an accurate model of neutrino interactions.  These models have to predict not only the signal and background populations that oscillation experiments see at near and far detectors, but they must also predict how the neutrino's energy which enters a nucleus gets transferred to energies of the particles that leave the nucleus after the neutrino interacts.  

Neutrinos can also serve as unique probes of the nucleus itself.  Electron scattering measurements have uncovered many mysteries of nuclear structure, but not all of those mysteries have been solved.  Neutrinos, because they sample the quarks and the nucleons in a nucleus differently from charged leptons, can provide new insight into the nuclear environment.  In addition, neutrinos can sample the axial vector component of the cross section, and not simply the vector component sampled by charged leptons.  

The effect of the nucleus on neutrino interactions is a relatively new area of study: the earliest neutrino interaction measurements aimed at understanding the bare interactions with protons and neutrons, and so experiments tried to focus primarily on hydrogen and deuterium targets.  However, since oscillation experiments rely on kton-scale far detectors, they are forced to use much more common materials such as carbon, water, argon, or iron.  The nuclear effects even on an element as light as carbon are predicted to be substantial and must be modeled by neutrino oscillation experiments~\cite{mosel1}.  

Because neutrino oscillation probabilities are a function of the inverse of the neutrino energy, the current and next generation accelerator-based neutrino experiments are focusing on neutrino energies of a few hundred MeV to a handful of GeV.  At these energies, there are several important interaction channels.  At the energies of T2K\cite{t2k1,t2k2} and NOvA\cite{nova} the Quasi-elastic process where there is only a muon and nucleons in the final state, is a large fraction of the signal population.   The next most important process is charged pion production:  for a Cerenkov detector experiment this can pose as a background since the pion goes undetected and compromises the neutrino energy reconstruction.  Even for a totally active detector like scintillator or a time projection chamber (TPC), charged pion production can still cause ambiguities in neutrino energy measurements because the pion can be absorbed in the nucleus before it ever reaches the active detector material.  Neutral pion production, while less probable than charged pion production, still contributes a background in electron neutrino appearance searches and must be well-simulated.  Finally, deep inelastic scattering events can also contribute neutral pions which may contaminate an electron neutrino appearance measurement. 

 These channels must be well understood not just for neutrinos but also for antineutrinos, as well as for electron flavored neutrinos and antineutrinos.  Future oscillation experiments such as DUNE~\cite{dune} depend on the ability to predict far detector signal (background) spectra at the 1\% (5\%) level, and as this report will summarize, the particle physics community is still at the level of measuring cross sections and making far detector predictions at the 7-10\% level.  

One of the challenges of making precise neutrino cross section measurements is that the incoming neutrino flux is not known perfectly.  Experiments have recently achieved flux uncertainties between 8\% to 10\% at the focusing peak by using results from hadron production cross section measurements, made either on thin or on replica targets of the same material as the beamline target.  T2K, for example, has made use of thin and thick target measurements made by the NA61/SHINE experiment~\cite{shine}, and MINERvA is currently incorporating hadron production measurements from the MIPP experiment~\cite{mipp} which measured production on a graphite target that was later used in the neutrino beamline.  One potential area for future improvement on flux predictions comes from neutrino electron scattering, one of the few neutrino interactions whose cross section is known to QED-like precision.  Currently MINERvA has a ~15\% constraint from a measurement of that process but at low statistics~\cite{jaewon}, for the higher intensity beams of the future this can provide a 7\% constraint.   If the flux uncertainties are at the 10\% level and the statistical uncertainties are much lower, many cross section analyses optimize the statistical power of the data by measuring the shapes of the differential cross sections, or sometimes the ratios of cross sections between different targets when available.  

Over the past year there have been a number of new results released on each of the channels listed above using several different target nuclei.  These results are often not in agreement with predictions extraolated from electron scattering measurements, or even from predictions anchored to deuterium or hydrogen measurements.  These new measurements are starting to give theorists the handles they need to improve the theoretical description of neutrino interactions, which will ultimately pave the way for precision oscillation measurements.  This report briefly summarizes recent results and points out where those results differ from the predictions from past experience.  One unifying feature of these recent measurements is that they are compared to the GENIE\cite{genie} event generator, which is used by several current neutrino oscillation experiments, including T2K, NOvA, MINERvA, and ArgoNeuT.  

\section{Charged Current Quasi-Elastic Scattering}

This is one of the most important processes to understand as it is the dominant signal mechanism for T2K and supplies a large fraction of events in the NOvA experiment.  The charged current quasi-elastic (CCQE) cross section is often anchored to measurements of the elastic scattering cross section from charged leptons, and then the axial vector part can be parameterized through an axial form factor $F_a(Q^2)$ which takes as an input parameter the axial mass $M_A$ that governs not only the cross section as a function of $Q^2$ but also the absolute cross section level.  Previous measurements from deuterium are consistent with an axial mass of $1.03\pm0.02$~GeV~\cite{bernard}, while the MiniBooNE experiment measured a level and $Q^2$ dependence on carbon that are more consistent with an axial mass of $1.35\pm0.17$~GeV~\cite{mbooneqe}.  At first the assumption was that this cross section increase was due to unpredicted nuclear effects, but then NOMAD measured cross sections on Carbon at much higher energies that were consistent with the deuterium measurements. 

Since the last NuPhys conference there have been several new quasi-elastic measurements released, each of which have different techniques of isolating the reaction, tuning a background prediction and subtracting backgrounds, and extracting a result:  either the differential cross sections as a function of $Q^2$, or an extraction of the axial mass $M_A$.  Unless otherwise indicated, $Q^2$ for these experiments is determined by using a quasi-elastic scattering hypothesis where the initial nucleon is at rest and by reconstructing the momentum transfer simply by measuring the outgoing muon's momentum and angle, and assuming conservation of total momentum and energy. 

The MINOS experiment extracted $M_A$ on an iron target using NuMI, a wide-band neutrino beam at Fermilab peaked at 3.5~GeV~\cite{numibeam}.   MINOS selects events by requiring candidates to have a recoil energy below 225~MeV, tunes to several  different sidebands to predict the resonance, transition, and Deep Inelastic Scattering events which fall in the signal region.  They then fit a background-subtracted $Q^2$ distributtion to extract $M_A$ and find that $M_A=1.23^{+0.13}_{-0.09} (fit) ^{+0.12}_{-0.15} (syst)$.  The dominant systematic uncertainties are due to the uncertainties in intranuclear scattering, the hadronic energy cut, the detector model, and uncertainties in the low $Q^2$ suppression that the sideband analysis suggests for $\Delta$ production~\cite{minosqe}.  This result is between both the MiniBooNE and the NOMAD measurements, although they are also in an intermediate energy and on iron rather than carbon.  

The MINERvA experiment extracted the neutrino and antineutrino CCQE cross sections on carbon, also using the NuMI beam.  MINERvA also isolates events by cutting on the recoil energy that is away from the vertex region of the neutrino interaction, and tunes the background level in each $Q^2$ bin by fitting as a function of recoil energy the signal and sideband region for that $Q^2$ region.  The MINERvA neutrino and antineutrino cross section results are in better agreement with a prediction using a deuterium-like axial mass rather than the one consistent with MiniBooNE, but in fact sees the best agreement with a model based on electron scattering nuclear effects, which enhances the transverse part of the cross section.  MINERvA does not include energy near the vertex in order not to bias the analysis to the presence of extra nucleons emitted near the vertex, and then after the signal candidates have been chosen, also compares that vertex energy with a prediction that does not have multi-nucleon correlations.  MINERvA sees that in neutrino mode, the vertex energy is consistent with the emission of an extra proton below 225MeV in energy 25$\pm$9\% of the time, while in antineutrino mode the vertex energy indicates no extra protons.  So both the vertex energy and the $Q^2$ distributions are consistent with a multi-nucleon hypothesis~\cite{minervaqe,minervaqeb}.  

MINERvA has also extracted a quasi-elastic cross section in the NuMI neutrino beam as a function of the proton kinematics:  in this case MINERvA can require two tracks originating from a common vertex where one track is consistent with a muon but need not be momentum analyzed.  Then by measureing the outgoing proton momentum MINERvA can calculate the momentum transfered to the nucleus using the same assumptions described above.  The striking conclusion from these data is that the model that best describes the proton kinematics is one that has a simple relativistic Fermi Gas and no additional multi-nucleon effects~\cite{walton}.  

The T2K experiment, running in a neutrino beam which on axis (off-axis) is peaked at 1.5~GeV (700MeV), has recently released two different CCQE measurements, both on plastic but at the two different energies accessible in its near detector suite.  The fine-grained scintillator central module of INGRID, the on axis near detector,  was used to identify both one and two track events, and to search for multi-nucleon correlations by determining the ratio of one track to two track CCQE events,  This analysis finds the CCQE cross sections in both samples are in agreement more with the SciBooNE and MiniBooNE measruements  than the NOMAD measurement which is at higher energy~\cite{t2k_ccqe_onaxis}.   The fine grained tracker located off the main neutrino beam axis has also released an absolute cross section measurement~\cite{t2k_ccqe_offaxis}, again one that is consistent with the MiniBooNE data, and which can be parameterized as $M_A=1.26+0.21-0.18$~GeV/c$^2$.  

Finally, ArgoNeuT, running in the NuMI on axis low energy beam isolated charged current events and has examined them in more detail by measuring the number of final state protons that are emitted in events with little other vertex activity.  In particular, they have seen a surprising number of events with tracks that are consistent with two protons that are being emitted back-to-back from the nucleus.  The presence of these pairs is another hint that there are likely to be correlated nucleon pairs in an argon nucleus~\cite{hammer}.  

In conclusion, the new data mostly point towards additional nuclear effects in the CCQE process that are not yet included in the standard framework for neutrino event generators.  These additional effects have been measured by looking not only at the $Q^2$ or neutrino energy distributions measured using muon kinematics and assuming a quasi-elastic hypothesis, but also by looking in more detail at the hadronic side of the interaction.  Clearly, the best models of this process will need to incorporate both the hadronic side and the leptonic side of these interactions.  

\section{Inclusive Pion Production}

Unlike the CCQE process, the previous measurements of inclusive charged current pion production on hydrogen and deuterium~\cite{anl_pion, bnl_pion} were not consistent and disagreed with eachother at the 20\% level and were challenging to incorporate in neutrino event generators.  In addition, the more recent measurements the outgoing pion spectrum by MiniBooNE~\cite{miniboone_chpion} seemed to indicate that the produced pions undergo substantially less final state interactions than a model based on pion beam measurements would predict ~\cite{olga}.  

A recent re-analysis has resolved the earlier discrepancies in the deuterium data, by evaluating the ratios of charged pion to CCQE cross sections in both expreiments (which are in agreement) and then normalizing to the world average CCQE cross section on deuterium~\cite{wilkinson}.  

The MINERvA experiment has recently released neutrino charged pion production differential cross sections on scintillator using the NuMI beam centered at 3.5~GeV~\cite{eberly}.  These results show that the shape of the pion momentum spectrum is more consistent with predictions that have detailed modeling of final state interactions for the pions, rather than predictions that do not incorporate the effect of the nucleus.  However the level seen by MINERvA is apprlximately 20\% lower than the level predicted by GENIE.  This is in stark contrast to the pion momentum distribution from MiniBooNE, where the level predicted by GENIE was observed but the shape of the pion distribution was considerably different.  In both experiments the dominant pion production mechanism was through the Delta resonance.  These results taken together will help improve the models of the energy dependence of final state interactions, although a joint interpretation implies that something unexpected is occurring~\cite{sobczyk_pion, mosel_pion, Yu:2014yja}.  Between the time of the NuPhys conference and this writing, MINERvA has also released a neutral pion production cross section measurement~\cite{trung}, which also shows agreement with predictions from models incorporating full treatments of final state interaction.  

\section{Coherent Pion Production} 

Charged current coherent pion production is a rare and poorly-understood neutrino interaction.  In this process a neutrino scatters off an entire nucleus coherently and produces a very forward-going pion and transfers little or no energy to the nucleus.  The neutral current analog is a background with large uncertainties for electron appearance oscillation measurements, and the charged current analog has until recently only been seen at high energy neutrino experiments but not at the 1~GeV experiments like MiniBooNE and K2K.  

The MINERvA experiment has recently measured charged current coherent pion production on carbon in both its neutrino and antineutrino beams~\cite{higuera}.  This high-statistics measurement is unique at these energies in that the coherent interactions can be isolated using a model-independent cut on the momentum transfered to the nucleus, rather than a cut on the the momentum transfered to the combined pion-nucleus system.  The only cuts on the event sample were to isolate two-track events with a muon and pion candidate, where there is a minimum amount of vertex activity near the event vertex.  Once the signal candidates were isolated and backgrounds subtracted, the differential cross sections were compared to predictions from GENIE and NEUT, one of the event generators used by T2K.  Both the angular and momentum distributions measured differed considerably from those predicted in the standard neutrino event generators.  

Argoneut has also recently measured coherent pion production on argon in the NuMI beamline for both neutrinos and antineutrinos.  The Argoneut detector was too small to enable a reconstruction of the momentum transfered to the nucleus, so the analysis used a boosted decision tree to determine the probability that any given event was coherent.  The input to that procedure included the angles of the pion and muon tracks, the opening angles between the tracks, the kinetic energy of the pion based on calorimetry, the muon momenum, and the average stopping power of the first third of the muon track.  The levels seen in both neutrino and antineutrino events are consistent with that seen by MINERvA~\cite{argoneut_coherent} and current generators anchored on the high energy measurements. 

This is the continuation of an interesting story and the next step will be to measure the charged current coherent cross section at comparable precision across several different nuclei.  

\section{Charged Current Inclusive Scattering Measurements}

Argoneut has recently measured neutrino and antineutrino differential charged current cross sections on argon as a function of muon scattering angle and momentum~\cite{argoneut_ccinc}, and has found reasonable agreement with predictions based on both the GENIE simulation and a NuWRO\cite{nuwro} model, where the flux assumed was given by that constrained by the MINOS near detector flux analysis which normalizes to the total charged current cross section on iron~\cite{minos_flux}.  The measurement was done in a beam made while the negative pions were being focused, so the average antineutrino energy was 3.6~GeV while the average neutrino energy was at 9.6~GeV, since the neutrinos came primarily from positive pions that went through the center of the focusing system where the magnetic field was negligible.  The total $\nu_\mu$ ($\bar\nu_\mu$ CC cross section is $\sigma / E_\nu = 0.66 ± 0.03 ± 0.08 (0.28 ± 0.01 ± 0.03) \times 10^{−38} cm^2/$GeV per isoscalar nucleon at the two energies listed above, where the first error is statistical and the second is systematic.

T2K has measured the electron neutrino charged current inclusive cross section using its off-axis near detector, where the electrons are required to start in the fine grained detector, and the TPC and electromagnetic calorimetry are used for additional particle identification.  The photon background is constrained by looking at events containing converted photons.  The largest systematic uncertainties are from the flux (12.9\%), statistics 8.7\% and detector response (8.4\%).  This represents the first $\nu_e$ cross section measured at 1~GeV in nearly 30 years~\cite{t2k_nue_ccinc}.  A recent study has shown that electron neutrino-muon neutrino cross section ratios may not be as well-known as one might naively predict from the value of the electron and muon masses due to radiative corrections and other theoretical uncertainties~\cite{mcfarland_day}, so more measurements of electron neutrino charged current cross sections will be extremely helpful in untangling these effects.

\section{Charged Current Cross Section measurements as a function of Nucleus}

There are a variety of inclusive charged current cross section measurements that have been made in the past, but the ratio of cross sections between different nuclei is only starting to be measured at the 10\% level of precision.  Charged lepton measurements have seen a rich structure in this ratio as a function of the momentum of the struck quark in the parton model, or $x$~\cite{emc}.   However, the various kinematic regions studied have different behavior, and although the very lowest $x$ behavior is thought to be caused by shadowing or antishadowing, the "EMC effect", between $x$ of 0.4 and 0.7 has yet to be understood.  Different models of nuclear structure predict different ratios for this important $x$ region, and those models at best describe deep inelastic scattering measurements, but not the resonance or quasi-elastic regime.  One recent model by CTEQ compares the measured NuTeV cross sections to a predicted deuterium neutrino cross section constructed from parton distribution functions, and finds that the NuTeV data on iron~\cite{nutev_iron} does not match the prediction that would come from charged lepton scattering~\cite{morfin_emc}.  

T2K has used its near detector located on the beamline axis, INGRID, to measure the flux-averaged cross sections on iron and scintillator (CH), and has also produced a more precise ratio of cross sections, at a mean energy of 1.51~GeV~\cite{kikawa}.   The absolute cross sections have been measured with sub per cent statistical precision, and the dominant systematic uncertainties come from the flux prediction and are at the 10\% level.  The ratio has been measured at $1.047 \pm 0.007(stat.) \pm 0.035(syst.)$, and is consistent with the predictions from both GENIE and NEUT, two independent neutrino event generators in use at T2K.  

MINERvA  has released a measurement of the ratio of inclusive charged current cross section ratios between lead, iron, and carbon compared to scintillator (CH) as a function of neutrino energy and $x$~\cite{tice}.  This measurement was done by collecting events in the neutrino beam that originated near the passive nuclear targets located in the upstream region of the MINERvA detector, and then subtracting off the scintillator background by extrapolating from events in the data that originated in the downstream scintillator region of the detector.  Although the cross section ratios agree with the simulation as a function of neutrino energy, there is a significant disagreement in the ratios as a function of $x$.  For the low $x$ region, there is a deficit compared to the simulation that grows with the size of the nucleus, and for the high $x$ region, there is an excess compared to the simulation, which again grows with the size of hte nucleus.  The simulation comes from GENIE and is informed by charged lepton cross section ratio measurements.  Clearly more work is needed to better model the effects of the nuclear environment in order to better model the data.  Given the energy range of MINERvA, the excess at high $x$ is predominantly due to elastic processes, while the low $x$ region is dominated by more inelastic processes.  

\section{Future Prospects} 

The field of neutrino interactions is soon going to have significantly more constraints coming from the MINERvA and T2K near detector data that is already in hand.  Since the time of the NuPhys conference MINERvA has released a measurement of antineutrino production of neutral pions, the analog to the neutrino production of charged pion production described above.  MINERvA also expects to release significantly more data on the CCQE process:  the cross section ratio between $\nu_e$ and $\nu_\mu$, the $\nu_\mu$ CCQE cross section as a function of transverse and longitudinal muon momentum, and $\nu_\mu$ cross section ratios between lead, iron, and graphite to scinillator. 

T2K should also have a sizeable sample of electron neutrino CCQE events, and with the advent of antineutrino running now for the oscillation program, a host of antineutrino cross section measurements are now possible.  There is also an intriguing plan to look at cross sections at several different angles using the INGRID near detector, which would allow for almost monochromatic neutrino energy beams to be "created" by comparing event samples at  different off axis locations.  

On the longer horizon the MicroBooNE experiment will soon be collecting neutrino data and can provide a new look at many different neutrino interaction channels at high statistics~\cite{microboone} on argon.  Also, MINERvA is currently running in the Medium Energy neutrino beam, collecting much larger sets of events than what has already been reported, at twice the energy of the low energy beam.  The increased statistics will enable not only measurements of the same channels listed above at higher energies but comparisons across different nuclei~\cite{minerva_me}. 

Finally, there are several new ideas on the horizon that could completely change the level of precision in this field.  One idea, NuPRISM~\cite{nuprism} seeks to take advantage of the two-body  kinematics of pion kinematics to create monochromatic neutrino energy beams, again by comparing events at different off axis locations but in a large water cerenkov detector.  There are also several ideas for measuring neutrino interactions on Argon:  one idea involves using a high pressure gaseous argon Time Project Chamber (TPC), and there are plans to use Liquid Argon TPC's both at 1GeV energies significantly higher statistics than MicroBooNE with the Short Baseline Near Detector~\cite{lar1nd}, and at the 3-8GeV region accessed in the NuMI beam using the CAPTAIN detector~\cite{captain}.

\section{Summary}


Neutrinos have a long history of puzzling scientists:  although we understand now that they have mass and can change from one state to another over time, we are still trying find out if the antiparticles of neutrinos change at the same rates as neutrinos.  If they do change at different rates, that could help explain why the universe if filled with only matter and no antimatter.  Looking for these neutrino anti-neutrino differences means that we have to improve our understanding of the way they interact in the nuclei that make up our detectors.  This article describes a new suite of measurements of interactions on new precise detectors, measurements that are made possible by using the new intense neutrino sources that are also used to measure neutrinos over time.  This new suite of measurements, and those that are just around the corner, will help build models of neutrino interactions and in the process give us a new way of looking of the nucleus.  

The measurements described in this article, when taken together, indicate that the environment that a neutrino sees inside a nucleus is not as simple as the environment that either an electron sees, or the environment that the neutrino sees when interacting with a simple proton or a proton-neutron pair.   The more interaction channels we can measure and the more different nuclei we can test the better we can understand both neutrinos themselves and the nuclei where they interact.

\bigskip
\section{Acknowledgments}

This work was supported by the Fermi National Accelerator Laboratory, 
which is operated by the Fermi Research Alliance, LLC, 
under US Department of Energy contract
No. DE-AC02-07CH11359.

\end{document}